\documentclass[aps,prl,twocolumn,superscriptaddress,letterpaper,showpacs]{revtex4-1}%
\usepackage{graphicx}
\usepackage{verbatim}
\usepackage{natbib}
\usepackage[draft]{hyperref}%
\usepackage{amsmath}%
\usepackage{amsfonts}%
\usepackage{amssymb}
\usepackage{dcolumn}
\usepackage{color}
\providecommand{\U}[1]{\protect\rule{.1in}{.1in}}
\begin{document}
\title{Shot noise limited nanomechanical detection and radiation pressure backaction from an electron beam}
\author{S. Pairis}
\author{F. Donatini}
\affiliation{Univ. Grenoble Alpes, CNRS, Grenoble INP, Institut N{\'e}el, F-38000 Grenoble, France}
\author{M. Hocevar}
\affiliation{Univ. Grenoble Alpes, CNRS, Grenoble INP, Institut N{\'e}el, F-38000 Grenoble, France}
\affiliation{CNRS, Inst. NEEL, "Nanophysique et semiconducteurs" group, 38000 Grenoble, France}
\author{D. Tumanov}
\affiliation{Univ. Grenoble Alpes, CNRS, Grenoble INP, Institut N{\'e}el, F-38000 Grenoble, France}
\affiliation{CNRS, Inst. NEEL, "Nanophysique et semiconducteurs" group, 38000 Grenoble, France}
\author{N. Vaish}
\affiliation{Univ. Grenoble Alpes, CNRS, Grenoble INP, Institut N{\'e}el, F-38000 Grenoble, France}
\affiliation{CNRS, Inst. NEEL, "Nanophysique et semiconducteurs" group, 38000 Grenoble, France}
\author{J. Claudon}
\affiliation{Univ. Grenoble Alpes, CEA, INAC, PHELIQS, “Nanophysique et semiconducteurs“ Group, F-38000 Grenoble, France
}
\author{J.-P. Poizat}
\affiliation{Univ. Grenoble Alpes, CNRS, Grenoble INP, Institut N{\'e}el, F-38000 Grenoble, France}
\affiliation{CNRS, Inst. NEEL, "Nanophysique et semiconducteurs" group, 38000 Grenoble, France}
\author{P. Verlot}
\email{pierre.verlot@univ-lyon1.fr}
\affiliation{{Universit{\'e} Claude Bernard Lyon 1, UCBL,}\\ Domaine Scientifique de La Doua, 69622 Villeurbanne, France}
\date{\today}

\begin{abstract}
Detecting nanomechanical motion has become an important challenge in Science and Technology. Recently, electromechanical coupling to focused electron beams has emerged as a promising method adapted to ultra-low scale systems. However the fundamental measurement processes associated with such complex interaction remain to be explored. Here we report highly sensitive detection of the Brownian motion of $\mu\mathrm{m}$-long semiconducting nanowires (InAs). The measurement imprecision is found to be set by the shot noise of the secondary electrons generated along the electromechanical interaction. By carefully analysing the nano-electromechanical dynamics, we demonstrate the existence of a radial backaction process which we identify as originating from the momentum exchange between the electron beam and the nanomechanical device, which is also known as radiation pressure.
\end{abstract}

\pacs{42.70.Qs, 43.40.Dx}
\maketitle

\paragraph{Introduction}
Nanomechanical devices are raising increasing interest both in Science and Technology \cite{cleland2013foundations}: Thanks to their reduced size and masses, these solid state systems are weakly impacted by decoherence mechanisms \cite{zurek2003decoherence}, with the additional asset to be repeatably (and almost infinitely) measurable. These unique properties provide nanomechanical resonators with an outstanding sensing potential which is being exploited in a variety of contexts, including quantum physics \cite{Arcizet2011,yeo2014strain}, ultra-sensitive force measurements  \cite{rugar2004single}, bio-sensing \cite{tamayo2013biosensors}, nanotribology \cite{bhushan2005nanotribology} and mass spectroscopy \cite{Chaste2012}. Decreasing the dimensions of nanomechanical resonators reinforces the challenge of detecting their mechanical motion, whose coupling to the measurement probe is generally diminished at lower scales. Concurrently, dynamical effects induced by the probe itself are generally enhanced, as exemplified in nano-optomechanical systems \cite{anetsberger2009near,eichenfield2009picogram} which have been introduced for that very purpose, with the perspective to study the fundamental processes in quantum macroscopic measurements \cite{aspelmeyer2014cavity}. In particular, electromechanical coupling to focused electron beams has been recently pointed out as a very promising alternative to optical schemes for nanomechanical systems with dimensions well below the diffraction limit \cite{Nigues2014a,tsioutsios2017real}. However the potential of this method remains largely unknown, essentially because of its strongly dissipative nature \cite{seiler1983secondary}.

In this Letter, we experimentally investigate the elementary physical processes associated with the electromechanical coupling between a focused electron beam and a nanomechanical resonator. We report ultra-sensitive detection of the Brownian motion of $\mu\mathrm{m}$-long InAs nanowires, with a sensitivity that can be as low as $S_{\mathrm{xx}}^{\mathrm{imp}}=(270\,\mathrm{fm}/\sqrt{\mathrm{Hz}})^2$, comparable or even better than the state-of-the-art for equivalent probe powers \cite{nichol2008displacement}. We demonstrate that the sensitivity is set by the gradient of secondary electron emission, with an imprecision originating from the shot noise of the emitted secondary electrons. We show that the geometry of our experiment enables to extract the fundamental component of the measurement backaction process and demonstrate the existence of a radial force associated with the nano-electromechanical measurement, which we identify as the radiation pressure force exerted by the electron beam on our nanomechanical structure. In a more general perspective, our work and methods show that motion correlations between the two orthogonal modes of our 2-dimensional resonator enable to reveal sensitive information on the origin of the measurement backaction force, which may be further extended to various physical contexts such as light momentum measurements \cite{beth1936mechanical,kippenberg2008cavity} and ultra-sensitive force microscopy \cite{de2017universal,rossi2016vectorial}.

\begin{figure}[h]
\includegraphics[width=\columnwidth]{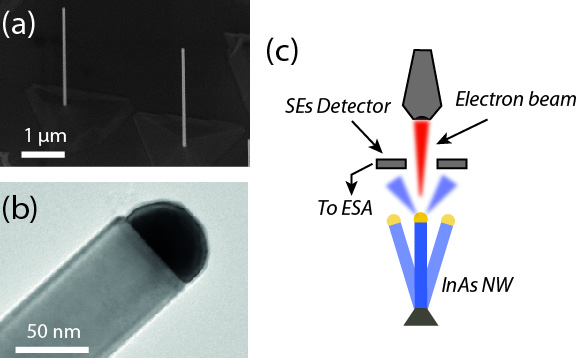} \centering
\caption{(a) Scanning Electron Micrograph showing two InAs nanowires similar to those used in this work ($\simeq 25^{\circ}$ tilted view). (b) Transmission Electron Microscopy (TEM) image of the upper part of a InAs nanowire. The dark hemisphere at the top of the nanowire is the gold catalyst. (c) Schematic depicting the principle of the experiment. A focused electron beam is sent on the nanowire, whose vibrations result in fluctuations of the secondary electron current. These fluctuations are monitored using a secondary electrons detector which is further connected to an electrical spectrum analyser (ESA).}%
\label{Fig1}%
\end{figure}

\begin{figure}[h]
\includegraphics[width=\columnwidth]{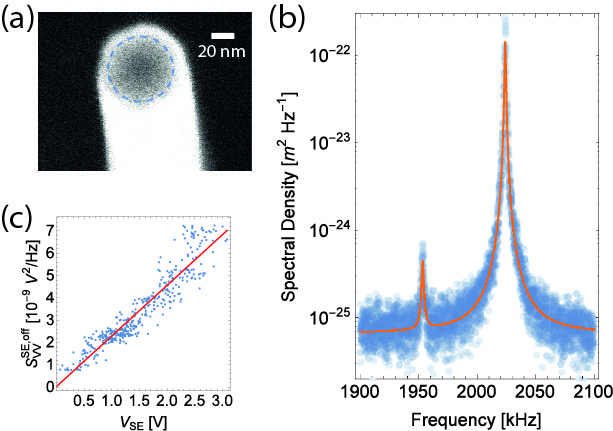} \centering
\caption{ (a) Magnified scanning electron micrograph showing an InAs NW similar to those used in this work (top view). The dashed line emphasizes the detection annulus, that is the region where the e-beam spot is positioned for measuring nanomechanical motion (see text). (b) Electromechanical fluctuations spectrum obtained with $\mathrm{NW}_1$. Two peaks are observed, corresponding to each of the two perpendicular vibrational directions. The experimental data (dots) are fitted using a double Lorentzian model (solid line), from which a mechanical resonance frequency $\Omega_{1}/2\pi\simeq2023.9\,\mathrm{kHz}$ and a mechanical quality factor $Q_1\simeq1752$ are found for the highly resolved peak. The shot noise limited measurement imprecision is determined from the detection background to the level of $S_{\mathrm{xx}}^{\mathrm{imp}}[\Omega\simeq\Omega_{1}]\simeq(270\,\mathrm{fm}/\sqrt{\mathrm{Hz}})^2$ with respect to that mode. (c) Evolution of the measurement noise background as a function of the average secondary electrons intensity. The red, solid line corresponds to a linear model, characteristic of the granular nature of SEs emission.}%
\label{Fig2}%
\end{figure}

\paragraph{Experimental setup}
The nanomechanical systems used in this work consist of as-grown InAs nanowires. We grew the nanowires perpendicularly on a (111)B InAs substrate by the vapour solid liquid method using 50 nm-gold catalysts in a molecular beam epitaxy setup. The nanowires were grown at a growth rate of 0.4 monolayers per second, a V/III beam equivalent pressure ratio of 50 and a substrate temperature of $420^{\circ}C$. The resulting nanowires are surmounted by a hemispherical gold droplet (see Fig. \ref{Fig1}(b)) and feature a wurtzite crystal structure with a limited number of stacking faults. The nanowires have lengths and diameters typically ranging from $4\,\mu\mathrm{m}-5.5\,\mu\mathrm{m}$ and $60\,\mathrm{nm}-80\,\mathrm{nm}$, respectively. Figure \ref{Fig1}(a) shows a tilted scanning electron micrograph of nanowires similar to those used in this work.  The results hereby reported have been obtained using three distinct samples referenced as $\mathrm{NW}_{j\in\{1,2,3\}}$ in the following.  The samples are mounted in a field emission scanning electron microscope operating with a probe current set to $I_{\mathrm{p}}=186\,\mathrm{pA}$ and an acceleration voltage $V=3\,\mathrm{kV}$. The 3D positioning stage hosting the sample is subsequently carefully aligned for matching the electron beam direction to the axis of the nanowires. Measuring the defocusing between the wafer plane and the edge of the nanowires enables to estimate a residual tilt angle $\alpha\simeq2.4^{\circ}$. Figure \ref{Fig2}(a) shows a typical scanning electron micrograph obtained in such conditions with our samples.

The nanomechanical motion of the nanowire is detected by setting the electron beam spot to a high-contrast region of the tip surface, on the external annulus delimiting the central, dark region (which will be recalled as "detection annulus" in the following, see Fig. \ref{Fig2}(a)). The vibrations of the nanowire result in fluctuations of the secondary electrons (SEs) current, which are directly monitored by connecting a low noise electrical spectrum analyser to the amplified SEs detector output \cite{Nigues2014a,tsioutsios2017real} (see Fig. \ref{Fig1}(c)). The electromechanical spectrum obtained with $\mathrm{NW}_1$ is shown in Fig. \ref{Fig2}(b). Two peaks are revealed, corresponding to each of the two vibrational directions of the nanowire. Remarkably, the highest noise peak is resolved with a signal-to-noise ratio exceeding $30\,\mathrm{dB}$ despite the relatively high mechanical frequency. To calibrate the electromechanical fluctuation spectrum, we assume the nanomechanical noise to be thermally driven - this hypothesis will be further confirmed -with effective mass $m_{1}=22\,\mathrm{fg}$ determined from the sample geometry, mechanical resonance frequency $\Omega_{1}/2\pi=2023.9\,\mathrm{kHz}$, and temperature $T\simeq 300\,\mathrm{K}$. The SEs fluctuations are subsequently converted into equivalent displacement by matching the SEs current variance with the thermal noise variance $(\Delta x_{\mathrm{th},1})^2=\frac{k_BT}{m_{1}\Omega_{1}^2}$. In particular, we find a displacement sensitivity $S_{\mathrm{xx}}^{\mathrm{imp}}\simeq (270\,\mathrm{fm}/\sqrt{\mathrm{Hz}})^2$, which even exceeds the performances of the most sensitive optical cavity-less detection schemes for free standing resonators with comparable dimensions \cite{nichol2008displacement}. The origin of the detection background is determined by measuring the evolution of the off-resonant spectral density of the SEs current $S_{\mathrm{II}}^{\mathrm{SE,off}}$ as a function of the average scattered current $\overline{I}_{\mathrm{SE}}$. We find a linear relationship, $S_{\mathrm{II}}^{\mathrm{SE,off}}\propto \overline{I}_{\mathrm{SE}}$, indicating that the SEs beam is shot noise limited (see  Fig. 2(c)).

\paragraph{Two-dimensional measurement \& sensitivity}
To further address the behaviour of the electromechanical coupling, we acquire SE's fluctuation spectra while browsing the e-beam spot position all around the edge of the dark central disk. Since we are detecting variations of the SE's emission rate, the corresponding intensity fluctuations $\delta I_{\mathrm{SE}}$ can be generally written as:

\begin{eqnarray}
\delta I_{\mathrm{SE}}(\mathbf{r_0},t)&\simeq&\nabla \overline{I}_{\mathrm{SE}}(\mathbf{r_0})\cdot\delta \mathbf{r}(t),\label{eq:1}
\end{eqnarray}

with $\nabla \overline{I}_{\mathrm{SE}}$ the SEs intensity gradient, $\mathbf{r_0}$ the average, two-dimensional position of the nanowire's tip in the horizontal plane and $\delta \mathbf{r}(t)$ the time dependent position variations, $\delta \mathbf{r}(t)=\delta x_1(t)\mathbf{e_1}+\delta x_2(t)\mathbf{e_2}$ ($\mathbf{e_{1,2}}$  the eigendirections of vibration and $\delta x_{1,2}(t)$ the associated displacements fluctuations). Because of the rotational symmetry (see Fig. \ref{Fig2} (a)), the  intensity gradient is radial, $\nabla \overline{I}_{\mathrm{SE}}=\left(\frac{\partial \overline{I}_{\mathrm{SE}}}{\partial r}\right)_{\mathbf{r_0}}\mathbf{e_r}$ where ($\mathbf{e_r}$,$\mathbf{e_\theta}$) is the polar base defined as $\mathbf{e_r}=\cos{\theta}\mathbf{e_1}+\sin{\theta}\mathbf{e_2}$ and  $\mathbf{e_\theta}=-\sin{\theta}\mathbf{e_1}+\cos{\theta}\mathbf{e_2}$). The SEs intensity fluctuations therefore reflect the combination of both displacements $\delta x_1(t)$ and $\delta x_2(t)$, weighted by their projection on the intensity gradient, $\delta I_{\mathrm{SE}}(\mathbf{r_0},t)=\left(\frac{\partial \overline{I}_{\mathrm{SE}}}{\partial r}\right)_{\mathbf{r_0}}\times\left(\cos{\theta} \delta x_1(t)+\sin{\theta} \delta x_2(t)\right)$, from which the expression of the electromechanical fluctuations spectrum (defined, in the limit of stationary fluctuations as $S_{\mathrm{II}}^{\mathrm{SE}}[\mathbf{r_0},\Omega]=\int_{-\infty}^{+\infty}\mathrm{d}t\mathrm{e}^{-i\Omega t}\langle\delta I_{\mathrm{SE}}(\mathbf{r_0},0)\delta I_{\mathrm{SE}}(\mathbf{r_0},t) \rangle$), can be inferred:
\begin{eqnarray}
S_{\mathrm{II}}^{\mathrm{SE}}[\mathbf{r_0},\Omega]=&&\nonumber\\
&\left(\frac{\partial \overline{I}_{\mathrm{SE}}}{\partial r}\right)_{\mathbf{r_0}}^2\times\left(\cos^2{\theta}S_{\mathrm{xx},1}^{r_0,\theta}[\Omega]+\sin^2{\theta}S_{\mathrm{xx},2}^{r_0,\theta}[\Omega]\right.&\nonumber\\
&\left.\,\,\,\,\,\,\,\,\,\,\,\,\,\,\,\,\,\,\,\,\,\,\,\,\,+\sin{2\theta}\mathrm{Re}\{\langle\delta x_1[\Omega]\delta x_2[-\Omega]\rangle_{r_0,\theta}\}\right)&\nonumber\\
 \label{eq:2}
\end{eqnarray}
\begin{figure}[h]
\includegraphics[width=\columnwidth]{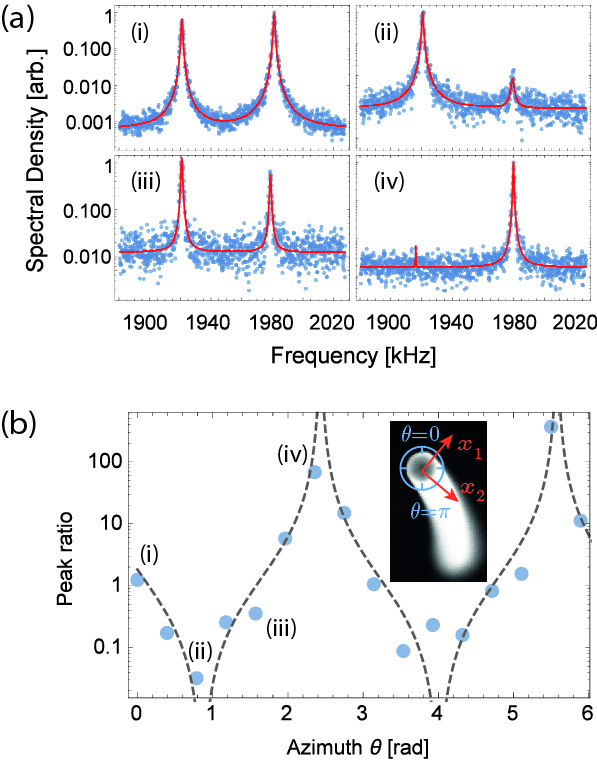} \centering
\caption{(a) Four electromechanical fluctuation spectra acquired at 4 distinct azimuthal positions of the circumference of the nanowire $\mathrm{NW}_2$ ($\theta=\pi/16,\,3\pi/16,\,5\pi/16$, and $7\pi/16$ from (i) to (iv), see Fig.\ref{Fig3}(b) for conventions). The balance of the two peaks is clearly changed, consistent with the radial direction of the secondary intensity gradient. (b) Ratio of the spectral amplitudes of the two peaks as a function of the azimuth. The experimental data points (dots) are fitted using a tangent squared model (dashed line), whose asymptotes enable to determine the direction of the eigenaxis of vibration. Inset: Scanning electron micrograph of the device showing the conventions used for the azimuth as well as the inferred direction of the vibrational axis.}%
\label{Fig3}%
\end{figure}
with $\Omega$ the Fourier frequency, and $S_{\mathrm{xx},1}^{r_0,\theta}$ (resp. $S_{\mathrm{xx},2}^{r_0,\theta}$) the displacement fluctuations spectra associated with $\delta x_1$ (resp. $\delta x_2$). Note that the superscript $r_0,\theta$ is to remind that the motion spectral density includes the contribution of a measurement backaction \emph{a priori}, which depends on the electromechanical coupling rate, and henceforth on the polar coordinates. The second line of  Eq. \ref{eq:2} represents $\theta$-dependent motion correlations between the two vibrational directions, which occur in presence of a common external driving source \cite{caniard2007observation}, resulting in strong spectral distortions compared to the uncorrelated bi-Lorentzian model (first line of Eq. \ref{eq:2}).

 Figure \ref{Fig3}(a) shows 4 electromechanical spectra acquired at 4 distinct azimuths (data acquired with $\mathrm{NW}_2$). The experimental data (dots) are fitted using a standard uncorrelated bi-Lorentzian model (solid lines, with $S_{\mathrm{xx},j}^{r_0,\theta}[\Omega]=\frac{S_{\mathrm{FF},j}^{r_0,\theta}}{m_j^2((\Omega_j^2-\Omega^2)^2+\Gamma_j^2\Omega^2)}$, $S_{\mathrm{FF},j}^\theta$ the white force spectral density driving nanomechanical motion in direction $j$). Since no deviation from this model was observed for any azimuth, we therefore conclude that the correlation term of Eq. \ref{eq:2} vanishes, $\langle\delta x_1[\Omega]\delta x_2[-\Omega]\rangle_\theta=0$. We subsequently compute the ratio of the peak values $r^2[\theta]=\tan^2{\theta}\times S_{\mathrm{xx},2}^{r_0,\theta}[\Omega_2]/S_{\mathrm{xx},1}^{r_0,\theta}[\Omega_1]$ (see Fig. \ref{Fig3}(b)). The experimental data (dots) are adjusted using a $\tan^2$ model (dashed line), from which we deduce that $S_{\mathrm{xx},2}^{r_0,\theta}[\Omega_2]\simeq S_{\mathrm{xx},1}^{r_0,\theta}[\Omega_1],\,\forall\theta$. Assuming equal effective masses in both vibrational directions, $m_2=m_1=m$, we conclude that $S_{\mathrm{FF},1}^{r_0,\theta}\simeq S_{\mathrm{FF},2}^{r_0,\theta},\,\forall\theta$. This implies that e-beam induced backaction fluctuations have a vanishingly small contribution to the nanomechanical dynamics, therefore establishing thermal noise as the dominant random source of motion.  

\paragraph{Backaction gradients} 
To further investigate the backaction processes associated with the electromechanical measurement, we now examine the effects produced by their gradients, which generally leave much stronger dynamical signatures than fluctuations at room temperature (as e.g. for dynamical backaction in optomechanics \cite{Arcizet2006,kippenberg2008cavity}). The backaction force is essentially the sum of two contributions of different nature, $F_{\mathrm{ba}}=F_{\mathrm{d}}+F_{\mathrm{q}}$. Here $F_{\mathrm{d}}$ denotes the contribution of dissipative mechanisms, which result from heating due to e-beam absorption. Previous work has shown that electrothermal actuation is the dominant dissipative mechanism with semiconducting nanomechanical devices \cite{Nigues2014a}: A fraction of the electrical energy carried by the incident electron beam is released as heat, yielding to deformations that are equivalent to nanomechanical motion in one invariable direction (imposed by the imperfect geometry of the nanowire), $\mathbf{F_{\mathrm{\mathbf{d}}}}(r,\theta)=F_{\mathrm{d}}(r,\theta)\cos{\theta_d}\,\mathbf{e_1}+F_{\mathrm{d}}(r,\theta)\sin{\theta_d}\,\mathbf{e_2}$, with $F_{\mathrm{d}}$ the modulus of the electrothermal force and $\theta_d$ the direction of the force in the basis $(\mathbf{e_1},\mathbf{e_2})$. In contrast, $F_{\mathrm{q}}$ denotes the measurement backaction force, resulting from the only action of measuring the system, independent from the experimental environment \cite{braginsky1995quantum}. We attribute this force to radiation pressure whereby the incident electrons are transferring part of their momentum to the nanowire in the radial direction, $\mathbf{F_{\mathrm{\mathbf{q}}}}(r,\theta)=F_{\mathrm{q}}(r,\theta)\mathbf{e_\mathrm{r}}$, with $F_{\mathrm{q}}$ the modulus of the radiation pressure force. 

The effect of force gradients is to modify the effective restoring force in both nanomechanical motion directions, resulting in a frequency shift $\delta\Omega_j^k=\frac{1}{2m\Omega_j}\frac{\partial F_{k,j}}{\partial x_j}$ ($k\in\{\mathrm{d,q}\}$, $j\in\{1,2\}$), with $F_{k,j}=\mathbf{F_{\mathrm{\mathbf{k}}}}\cdot\mathbf{e_j}$. Taking the above given general expression for $\mathbf{F_{\mathrm{\mathbf{d}}}}$ and $\mathbf{F_{\mathrm{\mathbf{q}}}}$ subsequently yields to:
\begin{eqnarray}
2m\Omega_1\delta\Omega_1^{\mathrm{d}}&=&\cos{\theta}\cos{\theta_d}\frac{\partial F_{\mathrm{d}}}{\partial r}-\frac{\sin{\theta}\cos{\theta_d}}{r}\frac{\partial F_{\mathrm{d}}}{\partial\theta}\nonumber\\
2m\Omega_2\delta\Omega_2^{\mathrm{d}}&=&\sin{\theta}\sin{\theta_d}\frac{\partial F_{\mathrm{d}}}{\partial r}+\frac{\sin{\theta_d}\cos{\theta}}{r}\frac{\partial F_{\mathrm{d}}}{\partial\theta}\nonumber\\
2m\Omega_1\delta\Omega_1^{\mathrm{q}}&=&\frac{\sin^2{\theta}}{r}F_{\mathrm{q}}(r,\theta)+\cos^2{\theta}\frac{\partial F_{\mathrm{q}}}{\partial r}-\frac{\sin{2\theta}}{2r}\frac{\partial F_{\mathrm{q}}}{\partial\theta}\nonumber\\
2m\Omega_2\delta\Omega_2^{\mathrm{q}}&=&\frac{\cos^2{\theta}}{r}F_{\mathrm{q}}(r,\theta)+\sin^2{\theta}\frac{\partial F_{\mathrm{q}}}{\partial r}+\frac{\sin{2\theta}}{2r}\frac{\partial F_{\mathrm{q}}}{\partial\theta}\nonumber\\
\label{eq:3}
\end{eqnarray}
In addition to the effects of backaction gradients, the mechanical resonance frequencies may be prominently affected by temperature-induced internal changes of the nanomechanical system \cite{gavartin2013stabilization,Gloppe2014a}, resulting in common mode frequency variations $\delta\Omega_1^{\mathrm{th}}(\theta)=\delta\Omega_2^{\mathrm{th}}(\theta)=\delta\Omega_{\mathrm{th}}(\theta)=\sum_k\frac{\partial\Omega_0}{\partial p_k}\frac{\partial p_k}{\partial T}\delta T(\theta)$, with $\delta T$ the temperature variation and $p_k\in\{R_0, L, Y_{\mathrm{InAs}}, \rho_{\mathrm{InAs}}\}$ the $k^{\mathrm{th}}$ parameter involved in the expression of the intrinsic mechanical resonance frequency $\Omega_0\simeq\left(\frac{0.6 \pi}{L}\right)^2\sqrt{\frac{\pi Y_{\mathrm{InAs}}R_0^2}{\rho_{\mathrm{InAs}}}}$ (with $R_0$ the radius of the nanowire, $Y_{\mathrm{InAs}}$ Young's modulus and $\rho_{\mathrm{InAs}}$ the mass density). In total, each mechanical resonance frequency shift generally expresses as the sum of three terms, $\delta\Omega_i=\delta\Omega_i^{\mathrm{d}}+\delta\Omega_i^{\mathrm{q}}+\delta\Omega_{\mathrm{th}}$.

\begin{figure}[h]
\includegraphics[width=\columnwidth]{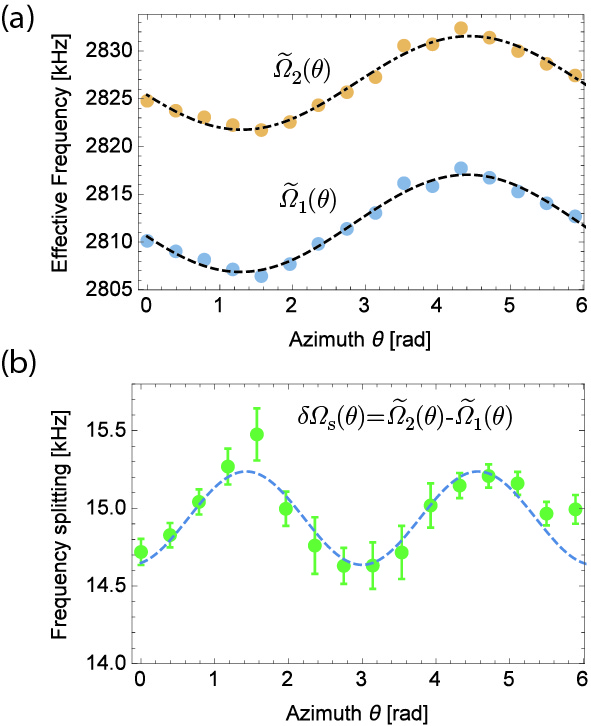} \centering
\caption{(a) Evolution of the effective mechanical resonance frequencies $\widetilde{\Omega}_1$ and $\widetilde{\Omega}_2$ as a function of the azimuth (data acquired with $\mathrm{NW}_3$). The experimental data (dots) have been acquired by browsing the internal circle of radius $r=20\,\mathrm{nm}$. The dashed lines are sinusoidal fits, resulting from the modulation of the absorbed energy as a function of the azimuth (see text). (b) Frequency splitting as a function of the azimuth (data acquired with $\mathrm{NW}_3$). The experimental data (dots) are fitted using a $\pi$-periodic sinusoidal model, characteristic of the (radial) radiation pressure backaction force (see text). The sine-wave amplitude $\Delta\Omega_s/2\pi=600\,\mathrm{Hz}$ yields a static backaction force value $F_{\mathrm{q}}\simeq 34\,\mathrm{fN}$.}%
\label{Fig4}%
\end{figure}

Figure \ref{Fig4}(a) shows the evolution of the mechanical resonance frequencies $\widetilde{\Omega}_i(\theta)=\Omega_i+\delta\Omega_i(\theta)$ of $\mathrm{NW}_3$ as a function of the azimuth of the electron beam spot on the detection annulus ($\Omega_i$ the intrinsic mechanical resonance frequency associated with mode $i$ ). To zeroth order, both frequencies are shifting from similar, sinusoidal amounts (dot-dashed and dashed lines). Such behaviour essentially reflects the contribution of temperature changes, $\delta\Omega_i(\theta)\simeq\delta\Omega_{\mathrm{th}}(\theta)$, since force gradients cannot generate identical $2\pi-$ periodic frequency shifts other than zero (see Supplementary Information).

The sinusoidal evolution of the temperature explains because of the small tilt angle $\alpha$ of the incident electron beam with respect to the top face of the nanowire, yielding to azimuth dependent energy deposition (see Supplementary Information). At room temperature, the coefficient of thermal expansion of $\mathrm{InAs}$ is on the order of $\alpha_{\mathrm{InAs}}\simeq 4.5\times10^{-6} \mathrm{K}^{-1}$ \cite{mikhailova1996indium}, negligible compared to the relative change of Young's modulus $\frac{1}{Y_{\mathrm{InAs}}}\frac{\partial Y_{\mathrm{InAs}}}{\partial T}\simeq 1.2\times 10^{-4}\mathrm{K}^{-1}$ \cite{burenkov1975elastic}, yielding to $\delta\Omega_{\mathrm{th}}(\theta)\simeq\frac{1}{2Y_{\mathrm{InAs}}}\frac{\partial Y_{\mathrm{InAs}}}{\partial T}\Omega_0\times\delta T(\theta)$. From the amplitude of the sine wave $\Delta\Omega_{\mathrm{th}}/2\pi=5.15\,\mathrm{kHz}$, it is possible to determine the total temperature variation over scanning the detection annulus $\Delta T=2\times 2Y_{\mathrm{InAs}}\left(\frac{\partial Y_{\mathrm{InAs}}}{\partial T}\right)^{-1}\times\frac{\Delta\Omega_{\mathrm{th}}}{\Omega_0}\simeq 58\,\mathrm{K}$, in reasonable agreement with simulations of energy absorption (see Supplementary Information). In a more general perspective, this result exemplifies how our electromechanical approach enables to perform thermal measurements \emph{in-situ}, and in particular to estimate e-beam induced heating in nanomechanical structures.

\paragraph{Radiation pressure contribution}
To complete our study, we examine the evolution of the frequency splitting $\delta\Omega_s(\theta)=\widetilde{\Omega}_2(\theta)-\widetilde{\Omega}_1(\theta)$, which enables to reject the common-mode frequency variations as a function of the azimuth. Because of the non-degenerate nature of the nanowire, the splitting reads $\delta\Omega_s(\theta)=\delta\Omega_{s,0}+\{\delta\Omega_2^{\mathrm{d}}(\theta)-\delta\Omega_1^{\mathrm{d}}(\theta)\}+\{\delta\Omega_2^{\mathrm{q}}(\theta)-\delta\Omega_1^{\mathrm{q}}(\theta)\}$, with $\delta\Omega_{s,0}/2\pi\simeq 15.2\,\mathrm{kHz}$ the bare fundamental resonance frequency splitting. From the above study of the thermal shifts, it is possible to show that dissipative backaction gradients do not contribute to the azimuthal variations of the frequency splitting (i.e. $\delta\Omega_2^{\mathrm{d}}(\theta)-\delta\Omega_1^{\mathrm{d}}(\theta)$ is $\theta$-independent, see Supplementary Information), any observed evolution being therefore necessarily attributed to the fundamental, radiation pressure component. Assuming rotational invariance ($\partial F_{\mathrm{q}}/\partial\theta=0$), the corresponding contribution can be further expressed as $2m\Omega_0\{\delta\Omega_2^{\mathrm{q}}(\theta)-\delta\Omega_1^{\mathrm{q}}(\theta)\}=\left(F_{\mathrm{q}}/R-\left(\partial F_{\mathrm{q}}/\partial r\right)_R\right)\cos{2\theta}$, which is a $\pi$-periodic sinusoidal function of the azimuth, noticeably. Fig. \ref{Fig4} (b) shows the experimentally obtained azimuthal evolution of the frequency splitting (dots). The dashed line is an offset, $\pi$-periodic sine wave fit, in excellent agreement with our model. Further assuming a quadratic form $F_{\mathrm{q}}(r)=\varphi''_{\mathrm{q}}r^2$ (which can be justified by the curved Secondary Electron intensity profile, see Fig. \ref{Fig2} (a)) enables to estimate the amplitude of the radiation pressure force exerted in the horizontal plane, $F_{\mathrm{q}}(R)=m\Omega_0\Delta\Omega_s R\simeq 34\,\mathrm{fN}$, with $\Delta\Omega_s/2\pi=600\,\mathrm{Hz}$ the amplitude of the sine wave fit. An interpretation of this value can be drawn by considering the associated uncertainty product $\Delta x^{\mathrm{imp}}\Delta p^{\mathrm{imp}}$, with $\Delta x_{\mathrm{imp}}^2=S_{\mathrm{xx}}^{\mathrm{imp}}\Delta\nu$ and $\Delta p_{\mathrm{imp}}^2=S_{\mathrm{pp}}^{\mathrm{imp}}\Delta\nu$ the variances of the position and momentum noises, respectively and $\Delta\nu$ the measurement bandwidth ($\Delta\nu=1/\Delta\tau$, $\Delta\tau$ the measurement integration time). The momentum and measurement force noises are related via the relation $\Delta F_{\mathrm{imp}}=\Delta p_{\mathrm{imp}}/\Delta\tau$, so that the uncertainty product can be rewritten as $\sqrt{S_{\mathrm{xx}}^{\mathrm{imp}}\times S_{\mathrm{FF}}^{\mathrm{imp}}}$, with $\sqrt{S_{\mathrm{FF}}^{\mathrm{imp}}}=F_{\mathrm{q}}/\sqrt{I_{\mathrm{p}}/e}\simeq 9.8\times 10^{-19}\,\mathrm{N}/\sqrt{\mathrm{Hz}}$, with $I_{\mathrm{p}}/e$ the incident electron flux ($e\simeq 1.6\times 10^{-19}\,\mathrm{C}$ the positron charge). We obtain $\sqrt{S_{\mathrm{xx}}^{\mathrm{imp}}\times S_{\mathrm{FF}}^{\mathrm{imp}}}\simeq 5300 \frac{\hbar}{2}$. While being much reduced compared to previous studies \cite{Nigues2014a}, this result indicates that the present electromechanical measurements operate far from the Heisenberg limit, for which a product of $\frac{\hbar}{2}$ is expected. This excess of imprecision may arise from two contributions. First, it is likely that we operate far from the Cram{\'e}r-Rao bound, that would correspond to the highest attainable displacement sensitivity \cite{helstrom1976quantum}: Indeed for symmetry reasons, the present study has been achieved by operating on the detection annulus, which is at the expense of a decreased secondary electron gradient, yielding to a much reduced displacement sensitivity. The second reason that may explain the observed imprecision excess is more fundamental and related to the massive nature of the electrons. As demonstrated above, the measurement imprecision is set by secondary electron shot noise, which depends on the secondary electron yield (SEY), that is the number of emitted secondary electrons per incident primary electron. The SEY is a function of the incident electrons velocity, which is determined by the acceleration voltage. For gold, the SEY peaks around $V\simeq 300\,\mathrm{V}$ \cite{petry1926secondary}, whereas we pump our systems using electrons that are more than three times faster ($V=3\,\mathrm{keV}$), resulting in a backaction noise excess.

\paragraph{Conclusion}
In conclusion, we have reported ultra-sensitive, shot-noise limited nano-electromechanical detection of very high frequency semiconducting nanowires. Placing ourselves in a radial detection geometry, we have been able to show that this technique comes with negligible backaction noise at room temperature. The measurement backaction manifests as frequency changes. By analysing the spectral behaviour as a function of the electron beam spot azimuth in the upper horizontal plane, we have shown that it is possible to isolate the contribution of the radiation pressure force gradient as opposed to dissipative backaction mechanisms. In a more fundamental context, our results show how introducing a second, "auxiliary" sensing dimension can be utilized for getting around dissipative backaction mechanisms, which may be further considered for improving measurement efficiencies down to the Heisenberg limit and beyond. Last, on a technological side, our work demonstrates the availability of novel measurement methods adapted to nano-material engineering and the exploration of electron-matter interaction processes.

\section{Acknowledgements}
We gratefully acknowledge M. Or{\'u}, D. Beznasiuk, E. Bellet-Amalric,  and J.-M. G{\'e}rard for fruitful and stimulating discussions. This work is supported by The French National Research Agency (projects NOFX2015 ANR-15-CE09-0016 and QDOT  ANR-16-CE09-0010).

\bibliography{bib}
\end{document}